\pdfoutput=1 
\newcommand{\state} {| \psi_{\rm PDC}\rangle }   		
\newcommand{\statebra} {\langle \psi_{\rm PDC} | }
\newcommand{\sket}{| \phi \rangle }   		
\newcommand{\sbra}{ \langle  \phi | }   		
\newcommand{\zerouno} { \! \! \! \!\! \phantom{\ket{0}}_1  \! \! \bra{0}}
\newcommand{\zerodue} { \! \! \! \!\! \phantom{\ket{0}}_2  \!\! \bra{0}}

\newcommand{\cappa}{ {\cal K} }   		

\newcommand{\w}{\vec{w}}			

\newcommand{\vxi}{\vec{\xi}}				

\newcommand{\Aptilde}{\tilde{{\cal A}}_p}				%
\newcommand{\Ap}{{\cal A}_p}				%

\newcommand{\x}{\vec{x}}

\newcommand{\sinc}{{\rm sinc}}
\newcommand{\q}{\vec{q}}

\newcommand{\nn}{\nonumber}
\newcommand{\bsub}{\begin{subequations}}
\newcommand{\esub}{\end{subequations}}
\newcommand{\beq}{\begin{equation}}
\newcommand{\eeq}{\end{equation}}
\newcommand{\beqa}{\begin{eqnarray}}
\newcommand{\eeqa}{\end{eqnarray}}
\newcommand{\beql}{\begin{subequations}\begin{eqnarray}}
\newcommand{\eeql}{\end{eqnarray}\end{subequations}}
\documentclass[pra,aps,amsmath, nofootinbib]{revtex4}
\usepackage{amsmath} 
\usepackage{graphicx}
\usepackage{float}   
\usepackage{verbatim}  
\usepackage{braket}
\begin{document}
\title{Dimensionality of the spatio-temporal entanglement of  PDC photon pairs}
\author{A.~Gatti$^{1,2}$,T. Corti$^2$, E.~Brambilla$^2$, }
\affiliation{$^1$ Istituto di Fotonica e Nanotecnologie del CNR, Piazza L. Da Vinci 32, Milano, Italy;  $^2$  
Dipartimento di Scienza e Alta Tecnologia dell' Universit\`a dell'Insubria, Via Valleggio 11,  Como, Italy}
\author{D.B.Horoshko}
\affiliation{Laboratoire PhLAM, Universit\'e de Lille 1, 59655 Villeneuve d'Ascq, France}
\affiliation{B.I.Stepanov Institute of Physics, NASB, Nezavisimosti Ave. 68, Minsk 220072 Belarus} 
\begin{abstract}
In this work the Schmidt number of the two-photon state generated by parametric-down conversion (PDC) is evaluated in the framework of a fully spatio-temporal model for PDC.   A comparison with the results obtained  in either purely spatial or purely temporal models shows that the degree of entanglement of the PDC state cannot be trivially reduced to the product of the Schmidt numbers obtained in models with lower dimensionality, unless the detected bandwidth is very narrow. This result is a consequence 
of the non-factorability of the state in the spatial and temporal degrees of freedoms of twin photons. 
In the limit of a broad pump beam, we provide a geometrical interpretation of the Schmidt number,  as the ratio between the volume of the phase matching region and of a correlation volume. 
\end{abstract}
\maketitle
\section*{Introduction}
\label{sec:intro}
The process of parametric down-conversion (PDC) occurring in a nonlinear crystal is a widely employed source of entangled photons, which are ubiquitous ingredients in modern quantum technologies. An appealing aspect of this source is the possibility of generating {\em high-dimensional entanglement}, both in the sense that entanglement is generated in various degrees of freedom of the photon pair (polarization, time-energy, position-momentum), and because spatial and temporal entanglement is realized in a high-dimensional Hilbert space, due to the naturally ultrabroad bandwidths of the process.  High dimensional entanglement is attracting more and more attention, because of its potential to increase the capacity and the security of quantum communication channels, and the precision of quantum metrological techniques \cite{hideas}. 
\par
In this context an obvious question concerns the effective dimensionality of the entanglement of the PDC state  (or alternatively the number of  entangled modes generated by the process),  usually quantified  by the so-called {\em Schmidt number} \cite{ekert1994,parker2000}. Traditional approaches typically concentrate on a single degree of freedom at a time, depending on the application considered. For example,  the dimensionality of the temporal entanglement has been  evaluated in various configurations, including spontaneous PDC \cite{law2000, mikhailova2008}, quantum frequency combs generated by a syncronously pumped optical parametric oscillator \cite{patera2010,pinel2012}, waveguided PDC \cite{avenhaus2009}. The degree of transverse spatial entanglement of PDC \cite{law2004,exter2006, dilorenzo2009} is of paramount importantance for assessing both the dimensionality of the orbital angular momentum entanglement  (see e.g. \cite{torres2003,molina-terriza2007,franke-arnold2008,dada2011,dilorenzo2009,dilorenzo2010}) and the resolution of quantum imaging techniques \cite{gatti2008}.  In these studies a net separation between the spatial and temporal degrees of feedom was often justified by the assumption of a narrow filtering in the neglected degree of freedom. 
\par. 
However,  as for many  nonlinear optical processes, PDC is ruled by phase matching, which imposes an angular dispersion relation linking the frequencies and the angles of emission of the generated photons in a non-factorable way. This implies a strong coupling between temporal and spatial degrees of freedom, which has been recently evidenciated by the so called {\em X-entanglement} \cite{gatti2009,caspani2010,brambilla2010, brambilla2012,ottavia2012}, a feature shown for example by type I bipotons in conditions close to collinear phase matching,  whose temporal delays and the transverse spatial displacements at the crystal exit face are  linked by a  proportionality relation (corresponding to an X-shape of the spatio-temporal correlation in any plane containing time and one transverse coordinate). The space-time coupling is often regarded as a negative feature, because it affects e.g. the purity of the purely spatial entanglement when temporal degrees of freedom are neglected \cite{osorio2008}. However,  it also represents a valuable resource for engineering the quantum state of biphotons, since the spatial degrees of freedom can be used to tailor the temporal entanglement \cite{gatti2009,caspani2010} in order e.g. to realize an ultrabroadband temporally entangled state \cite{ottavia2012}. 
\par
In this work we adopt a fully spatio-temporal model of PDC, in order to investigate the effect of the non-factorability of the state in space and time  on the dimensionality of the biphoton entanglement. We shall concentrate on the evaluation of the spatio-temporal  Schmidt number of PDC entanglement in the ultra-low gain regime of PDC \footnote{ The Schmidt decomposition of the state in the full spatio temporal domain is very hard to achieve(even numerically), due to the hyperbolic geometry of the phase matching relations.  However, the Schmidt number can be computed even without performing the decomposition.} 
Based on a general formula for the Schmidt number, which involves integrals in 12 and 6 dimensions, we both obtain numerical evaluations and  analytical results, the latter being valid when the profile  of the pump driving the process is broad enough.  
In the same limit, we introduce a useful geometric interpretation of our results, which shows that the Schmidt number quantifying entanglement is basically the ratio between the volume of the region where phase matching efficiently occurs and a correlation volume, thus being proportional  to the number of spatio-temporal correlated modes. 
\par
The same methods of analysis are used to calculate the Schmidt number in a purely 2D spatial model and in a purely 1D temporal model of PDC. An important result that will be demonstrated is that the Schmidt dimensionality of the 3D model of PDC cannot be trivially reduced to the product of the Schmidt numbers in the models of lower dimensionality. 
As a matter of fact,  the Schmidt dimensionality of the full PDC state is by far larger than what would be expected from an approach  that considers space and time as separable degrees of freedoms, showing that  an enormous number of entangled modes is available for the down-converted light.  This result is again a consequence of the intrinsic  non-factorability of the state in its spatial and temporal degrees of freedom, and shows that using a full spatio-temporal  model for describing PDC is essential in order to correctly quantify  the degree  of entanglement of the state. 
\par
The results presented in this work extend and complement  those presented in  \cite{Horoshko2012}, where a 2D spatio-temporal model (1 temporal dimension +1 spatial dimension) for PDC has been investigated. 
\section{Starting point } 
\label{sec:general}
In \cite{gatti2009, caspani2010, gatti2003} the spatio-temporal quantum properties of the PDC light were described by studying the evolution of the quantum field operators throught the nonlinear crystal, and deriving input-output relations linking the operators at the crystal output face with those at the entrance face. Here we consider the equivalent state-formalism, in which the state evolves from the input to the output face of the crystal. We focus on type I PDC, in the regime of ultra-low gain where the probability of generating a photon pair in each spatio-temporal mode is small (more precisely, the probability of generating more than one photon pair in each mode is negligible). 
The output biphoton quantum state can thus be written as a generic superposition of the vacuum state  $|0\rangle$ and of a state with two photons generated in all possible spatio-temporal modes: 
\beq
\state = |0\rangle  +  \int d\w_1 \int d\w_2 \, {\cal C}  (\w_1, \w_2) 
A^\dagger (\w_1) A^\dagger (\w_2) |0\rangle 
\label{state0}
\eeq
where $A$ is the quantum field operators for the down-converted field, and $\w$ indicates the full 3-D spatio-temporal Fourier coordinate with the short-hand notation 
\beq 
\w= (\q, \Omega) \, ,
\eeq
where $\q$ is the transverse component of the photonic wave-vector $\q= (q_x, q_y)$, with respect to the mean propagation direction $z$ of the pump field, and $\Omega= \omega -\omega_p/2 $ is the temporal frequency  shift from the central  frequency of the PDC emission. The coordinate in the direct transverse space-time domain will be denoted by
\beq 
\vxi= (\x, t) \, ,
\eeq
where $\x= (x,y)$ is the spatial coordinate spanning the transverse plane at the crystal exit face, and t is time, with  the convention
\beq
\w \cdot \vxi = \q\cdot \vxi - \Omega t \; .
\eeq
$ {\cal C} (\vec{w_1},\vec{w_2})$ in Eq.\eqref{state0} is the probability amplitude of generating a photon pair in the spatio temporal modes  $\vec{w}_1$ and $\vec{w}_2$, and can be determined by exploiting the equivalence with the field formalism developed in \cite{gatti2009, caspani2010, gatti2003}. In these references the biphoton amplitude was calculated in terms of the field-field correlation at the crystal output face 
\beqa
\psi (\w_1, \w_2) &=& \langle A (\w_1, l_c) A (\w_2,l_c) \rangle  \label{psidef} \\
  &=&  \frac{g} { (2 \pi)^{\frac{3}{2}} } 
\Aptilde  (\w_1 + \w_2) \, \sinc \frac{\Delta (\w_1,\w_2) }{2} \,  e^{ i \frac {\Delta (\w_1,\w_2)}{2}   }
\label{psi} 
\eeqa
where 
\begin {description}
\item - $A(\vec{w}, l_c) $ is the output field operator,  and the expectation in Eq.\eqref{psidef} is taken on the input vacuum state; 
\item{ -$g$} is the dimensionless gain parameter, proportional to the pump peak amplitude, the crystal length and the nonliner susceptibility; 
\item {-$\Aptilde$} is the Fourier transform of the pump beam profile at the crystal exit face: 
\beq
\Aptilde (\w) := \int  \frac{ d \vxi} { (2 \pi)^{3/2} }   \Ap (\vxi)    e^{-i \vxi\cdot \w} \: ,
\eeq
where normalization is such that $\Ap (\vxi =0) =1$; 
\item {-$\Delta$} is the phase matching function, which accounts for the conservation of longitudinal momentum in the microscopic PDC process 
\beq
\Delta (\w_1,\w_2) = \left[ k_{sz} (\w_1) + k_{sz} (\w_2) - k_{pz} (\w_1 +\w_2)\right] \, l_c
\label{Delta}
\eeq
$k_{sz}$ being the longitudinal component of the (ordinary) signal wave vector, $k_{pz}$ the analogous quantity for the (extraordinary) pump, and $l_c$ the crystal length. 
\end{description}
We remark that the right hand side of Eq.\eqref{psi} is the first order term in the  parameter $g  $  of a perturbative expansion of the full solution of the propagation equation of field operators in the nonlinear crystal, so that expression \eqref{psi} is valid only in the very low gain regime $g \ll 1 $.  Similarly, as it is well-known, the right hand side of Eq.\eqref{state0} 
shows the zeroth and first order terms in $g$ of a perturbative expansion of the full PDC state. 
\par
By using the equivalence between the two formalisms, and by calculating the field correlation on the generic output state 
\eqref{state0} we also obtain: 
\beqa
\psi (\w_1, \w_2) &=& \statebra A (\w_1, 0) A (\w_2,0) \state  \nn \\
&=& {\cal C} (\w_1, \w_2) + {\cal C} (\w_2,\w_1) = 2 {\cal C} (\w_1,\w_2) \, .
\eeqa
where we used the symmetry properties of the state. 
Thus the two-photon state has the well known form 
\beq
\state = |0\rangle  + \frac{1} {2} \int d\w_1 \int d\w_2 \, \psi (\w_1, \w_2) 
A^\dagger (\w_1) A^\dagger (\w_2) |0\rangle 
\label{state1}
\eeq
with the biphoton amplitude $\psi$ being given by \eqref{psi}.  
\par
Apart from the biphoton amplitude, the other quantity of interest is the coherence function of the signal field, which after long but simple calculations can be derived from \eqref{state1} as:   
\beqa
G(\w, \w') &:=& \statebra    A^{\dagger} (\w)   A (\w') \state  \label {G} \\
 &=& \int d \w_2 \psi^* (\w, \w_2) \psi (\w', \w_2 )  
\eeqa
From this equation, the  total number of PDC photons is obtained as  
\beqa
N  &=& \int d\w_1  \statebra   A^{\dagger} (\w_1)   A (\w_1) \state \nn \\
&=&  \int d\w_1 \int d\w_2 \left| \psi(\w_1, \w_2) \right|^2
\label{N}
\eeqa
\par
In type I PDC a signal and an idler fields are in principle not distinguishable.
A bipartition of the system can be introduced in various ways, 
for example by sorting photons depending on their propagation directions (positive or negative with respect to any transverse axis) \cite{brambilla2012}, which,  however, is a good bipartion  only in the case of  a very broad pump waist, when twin photons are always created with symmetric tranverse wave-vectors $\pm \q$. In the ultra-low gain regime where at most a single photon pair at a time is detected, a very general bipartition is obtained  by simply considering the two output modes of a symmetric beam-splitter
\beqa
A_1(\w) &=& \frac{1}{\sqrt 2} \left[ A(w) + i a_v(\w)\right]  \nn \\
 A_2(\w) &=& \frac{1}{\sqrt 2} \left[ i A(w) +  a_v(\w)\right] 
\label{bs} 
\eeqa
where $a_v$ is a vacuum field operator.
By substituting in the state \eqref{state1} the inverse of \eqref{bs} one gets two terms that describe creation of two  photons
into  each of the output modes 1 and 2 of the beam-splitter, and a term that creates one photon in mode 1 and one photon in mode 2. 
Following the literature treating the degree of entanglement in PDC \cite{law2000,mikhailova2008,law2004,exter2006}, we shall consider rather than the full PDC, the state vector conditioned to the measurement of a photon pair (the vacuum and two-photon terms are dropped). 
Assuming that detectors are placed at the two output modes,  and coincidences are detected,   the state   conditioned to the appearence of a coincidence takes the form (a part from global normalization factors) 
\beq
\sket =  \int d\w_1 \int d\w_2 \, \psi (\w_1, \w_2) 
A_1^\dagger (\w_1) A_2^\dagger (\w_2) |0\rangle_1 |0\rangle_2 
\label{state2}
\eeq
The degree of entenglement of such a  conditional  state has been investigated in previous literature in the purely temporal \cite{law2000, mikhailova2008}  or purely spatial \cite{law2004, exter2006,dilorenzo2009} domains. 
\section{The Schmidt number of PDC  entanglement: integral formula} 
A good quantifier of the degree of entanglement for continuous variable pure states is the so-called Schmidt number, 
defined as the inverse of the purity of the state of each separate subsystem
\beq
\cappa= \frac{1}{ {\mathrm Tr} \{\rho_1^2\} }
\eeq
where $\rho_1$ is the reduced density matrix of the subsistem 1. In connection with the Schmidt decomposition of the PDC conditional state,
the Schmidt number is recognized to give an estimate of the number of Schmidt modes participating in the entanglement, i.e. of the effective dimensionality of the entanglement \cite{exter2006}. 

We will derive an integral formula for the Schmidt number in the case of the conditional state \eqref{state2}, similar to what obtained in Refs.\cite{mikhailova2008},  \cite{dilorenzo2009}.  
First of all the state \eqref{state2} is not normalized, 
\beq
\langle \phi \sket  = \int d\w_1 \int d\w_2 \left| \psi(\w_1, \w_2) \right|^2 =N 
\label{enne}
\eeq
From the system conditional density matrix  
\begin{equation}
\rho= \frac{ \sket \sbra} {\langle \phi \sket} 
\label{rho} 
\end{equation} 
the reduced density matrix of the subsystem 1 can be calculated (Appendix \ref{A2})  as
\beqa
\rho_1 &=& {\mathrm Tr}_2 \{ \rho\} \nn \\
&=& \frac{1}{N} \int d\w_1 \int d\w_1' G(\w_1', \w_1) A^{\dagger}_1 (\w_1) \ket{0}_1 \zerouno A_1 (\w_1') \, .
\eeqa
Notice that in the limit where the coherence function becomes a Dirac-delta - i.e in the limit of a monochromatic plane-wave pump- the reduced density matrix becomes a sum of projectors onto 1-photon states. 
\par
Next we calculate the purity of such a reduced state:
\beq
{\mathrm Tr}_1 \{ \rho_1^2 \} = \frac{1}{N^2} \left[ \int d\w_1 \int d\w_1' \left|G(\w_1, \w_1')\right|^2 \right]
\eeq
An integral formula for the Schmidt number can be threfore written as: 
\beq
\cappa = \frac{N^2} {B} 
\label{kintegral}
\eeq
where
\beqa
B &=& \int d\w_1 \int  d\w_1' \left|G(\w_1, \w_1')\right|^2 \\
 &=& \int d\w_1 \int d\w_2 \int d\w_1' \int d\w_2' \left[  \psi (\w_1,\w_2) \psi (\w_1',\w_2')\right. \nn\\
& &  \left. \psi^* (\w_1,\w_2') \psi^* (\w_1',\w_2)\right]
\label{k2} 
\eeqa
and $N$ is given by Eq.\eqref{enne}. 
\section{The nearly plane-wave pump approximation }
\label{sec:NPWPA} 
In order to evaluate the Schmidt number of the two-photon state from formula \eqref{kintegral} we face the problem of calculating the $6$-dimensional and  $12$-dimensional integrals involved in the calculation of $N$ and $B$, respectively . These integrations can be numerically performed, but in the following we will provide more trasparent results based on the approximation of a broad enough pump profile. 
\par
Let us come back to the expression\eqref{psi} for the biphoton amplitude that we rewrite as 
\beqa
\psi' (\w_1, \w_2) &=&
\Aptilde  (\w_1 + \w_2)  V(w_1, w_2) \label{psip} \\
V(w_1, w_2) &=&
\sinc \frac{\Delta (\w_1,\w_2) }{2} \,  e^{ i \frac {\Delta (\w_1,\w_2)}{2}   }
\label{V} 
\eeqa
where $\Delta$ is the phase matching function defined in Eq. \eqref{Delta}, and we got rid of the  costant $g/(2\pi)^{3/2}$ that factors out in the ratio $\cappa =N^2/B$. We now introduce the pump spectral coordinates $\w_p =\w_1+\w_2:= (\q_p, \Omega_p)$. Provided that  $\sigma_p$ is the transverse waist of the pump beam at the output crystal face, and $\tau_p$ its duration, the pump Fourier tranform $\Aptilde$ dies out on the scales   
$\delta q_p= 2/\sigma_p$, $\delta \Omega_p= 2/\tau_p$.  This claim is  exactly true for a Gaussian pump profile
\beqa
{\cal A}_p(\x,t) &=& e^{-x^2/\sigma_p^2} e^{-t^2/\tau_p^2} \label{pumpxi} \\
\Aptilde(\q_p,\Omega_p) &=&  \frac{ \sigma_p^2 \tau_p}{ 2^{3/2}}    e^{-q_p^2\sigma^2/4 } e^{-\Omega_p^2 \tau_p^2/4} 
\label{pumpspectrum}
\eeqa
The function $V$ is strongly peaked along the curve where phase matching takes place. As elaborated in detail in the Appendix B of \cite{caspani2010},  for a broad enough pump the  variation of this function with respect to the pump spectral coordinates can be neglected. In other words, $V(w_1, -\w_1 +\w_p) $ does not vary significantly with $\w_p$ on the scale over which the pump Fourier profile dies out: 
\beqa
\Aptilde (\w_p)V(\w_1, -w_1 +\w_p ) &\approx& \Aptilde (\w_p) V(\w_1, -\w_1) \nn \\
&: =& \Aptilde (\w_p) V(\w_1)  
\label{NPWPA} 
\eeqa
We call this approximation nearly plane wave pump (NPWP) approximation. It is based on making a Taylor expansion of $V$ in power series of the pump variables $\w_p$, and on finding the conditions under which the first order terms of the expansion can be neglected with respect to the zeroth order term \cite{caspani2010}. These conditions can be summarized as:
\beqa
	\tau_p  &\gg& \tau_{GVM} =|\frac{l_c}{v_{gs}}-\frac{l_c}{  v_{gp}   } | \label{taup}  \\
	\sigma_p &\gg& l_{\mathrm walk-off}=|l_c \frac{\partial k_p}{\partial q_x}|. \label{sigmap} 
\eeqa
Here $\tau_{GVM}$ is the maximum  delay time between the signal and the pump wave in crossing the nonlinear crystal, due to the mismatch  between the  group velocities $v_{gs}, v_{gp}   $  of the ordinary signal and  extraordinary pump.  $l_{\mathrm walk-off}$ is the maximum lateral walk-off between the two waves, associated to tilt of the Poynting vectors   
In the example of a $4$mm BBO crystal, pumped at a wavelenght $\lambda_p =527$nm, we have   $\tau_{GVM} \approx 500$fs,   $l_{\mathrm walk-off} \approx 250\mu$m, so that the NPWPA is within the reach of practical experimental generation of PDC photon pairs. 
\par
\begin{figure}[h]
\centering
{\scalebox{.6}{\includegraphics*{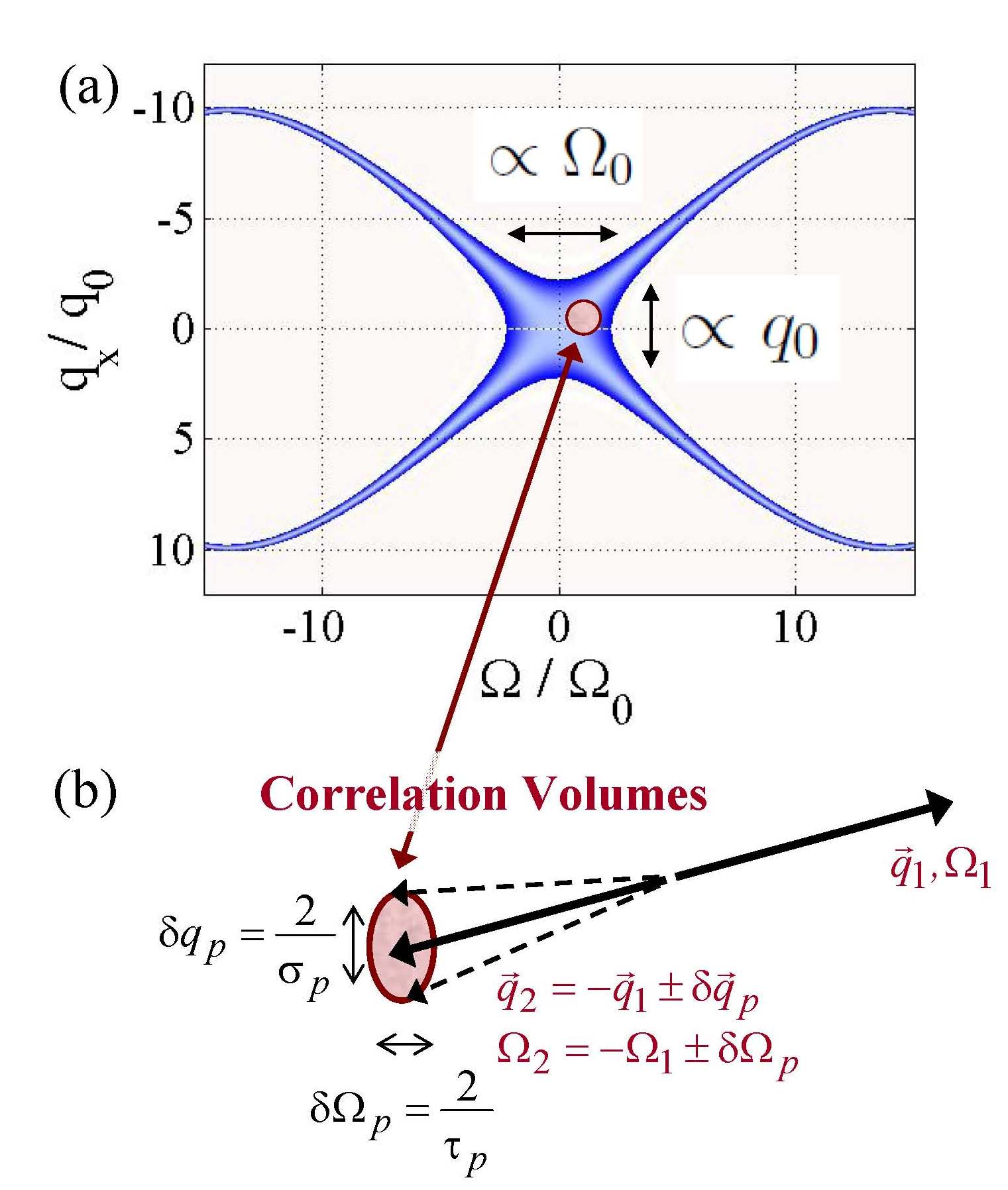}}}
\caption{ Geometrical interpretation of the Schmidt number. The figure plots  a) the phase matching region in the $(q_x, \Omega)$ plane, here defined by $|V(\vec{q},\Omega)|^2 > 0.1$ (the full 3D volume has a biconical shape),  and b)  the correlation  volumes, which in the 3D picture would be Gaussian bullets of size determined  by the spectral estension $\delta q_p^2 \delta \Omega_p$ of the pump.  Case of a 4 mm type I BBO crystal,  pumped at $\lambda_0=527$nm for collinear phase matching. }
\label{fig_spectrum}
\end{figure}

The use of this limit simplifies remarkably the expression \eqref{kintegral} of the Schmidt number. As reported in detail in the Appendix \ref{AppendixB}, the integral formula \eqref{kintegral} takes the form 
\begin{equation}
\mathcal{K}=\frac{N^2}{B} \to
\frac{  \left[ \int{d\vec{\xi}_p}|\mathcal{A}_p(\vec{\xi}_p)|^2 \right]^2}
{\int{d\vec{\xi}_p}|\mathcal{A}_p(\vec{\xi}_p)|^4} 
\; 
\frac{ \left[ \int{d\vec{w}}|V(\vec{w})|^2 \right]^2}
{(2\pi)^3   \, \int{d\vec{w}}|V(\vec{w})|^4} \, . 
\label{k2PW}
\end{equation}
The  integrals involving the pump coordinates can now  be readily performed. By assuming a Gaussian pump profile as in \eqref{pumpxi}, we easily obtain: 
\begin{equation}
	\cappa =
\frac{\int{d\vec{w}}|V(\vec{w})|^2}{\int{d\vec{w}}|V(\vec{w})|^4}\, 
 \frac{\int{d \vec{w}}\;\text{sinc}^2 \frac{\Delta(\vec{w})}{2}}{ \pi^{\frac{3}{2}}    \frac{4}{\sigma_p^2}\frac{2}{\tau_p}  }.
	\label{k3PW}
\end{equation}
As we shall see in the following,  under rather general conditions \footnote{this conditions amount to requiring that phase matching occurs within the spectral region considered, a conterexample being provided in Section\ref{sec_noncoll}} the first term at r.h.s. of Eq.\eqref{k3PW}  is a purely numerical factor, namely 
\begin{equation}
	\frac{\int{d \vec{w}\;|V(\vec{w})|^2}}{\int{d \vec{w}\;|V(\vec{w})|^4}}\;\approx\;\frac{3}{2}.
\end{equation}.
Thanks to this circumstance,  the  result of Eq.\eqref{k3PW} has a transparent  geometrical interpretation. The term 
$\int{d \vec{w}}\;\text{sinc}^2\frac{\Delta(\vec{w})}{2}$  can be interpreted as the volume of the region in the $ (\q, \Omega)$ 3D space where phase matching occurs, since the $\mathrm{sinc}^2$ function has a sharp maximum where $\Delta (\q, \Omega) =0$ (see Fig.\ref{fig_spectrum}). This corresponds to the portion of the $ (\q, \Omega)$ space where the probability of photon-pair production is not negligible. \\
On the other side the quantity at denominator is (a part from numerical factors) the spectral volume of the pump $\delta q_p^2 \delta \Omega_p =\frac{4}{\sigma_p^2} \frac{2}{\tau_p}$. This quantity defines the {\bf correlation volume}, i.e. the size the of the correlated modes, because the expression \eqref{psip} tells us that   in the NPWPA, the width of the biphoton correlation as a function of $\vec{w}_1+\vec{w}_2$  is determined by the pump Fourier profile. Thus $\delta q_p^2 \delta \Omega_p$  represents the uncertainty in the determination of the transverse wave vector $\q_2$ and frequency $\Omega_2$ of a photon, once the transverse wave vector $\q_1$ and frequency $\Omega_1$ of its twin have been determined. 
Therefore
$\cappa$ is on the order of the ratio:
\begin{equation}
	\cappa \; \propto \; \frac{\text{volume of the phase matching region}}{\text{correlation volume}}
	\label{intgeom} 
\end{equation}
and can be interpreted as the number of correlated modes partecipationg to the state. The formula \eqref{intgeom} gives us a simple geometric interpretation of the Schmidt number,  which will be very useful in understanding some results.
\section{Schmidt number of 3D X-entangled biphotons }
\label{Sec:3D}
\begin{figure}[h]
\begin{center}
\includegraphics[scale=0.5]{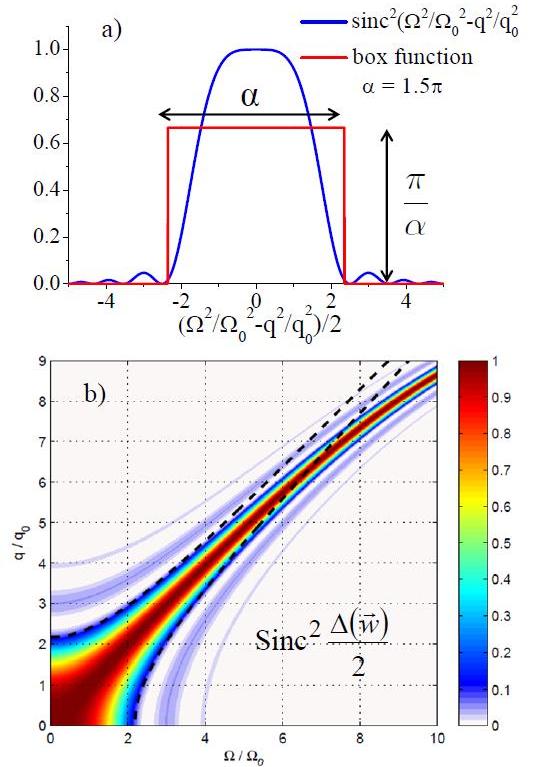}
\caption{a) Box function approximation (\ref{sincapprox}) of the $\text{sinc}^2$ function. b) Comparison between the true phase matching function $|V(\vec{q},\Omega)|^2$ and its box function approximation (boundaries of the box function are shown by  dashed lines). Collinear phase matching case ($\Delta_0 l_c =0$, $\theta_p = 22.934^o$),  $l_c=4$mm.  }
\label{figbox}
\end{center}
\end{figure}

We can proceed further,  and find an analytical approximation of  the Schmidt number result of Eq.\eqref{k3PW} 
in the NPWPA. To this end, we need to  calculate integrals 
over the phase matching curves of the form $\int{d \vec{w}}|V(\vec{w})|^2$, $\int{d \vec{w}}|V(\vec{w})|^4$. Our calculations are based on the use of two further approximations (in additions to the NPWPA): 
\\
i) The first approximation consists in a quadratic expansion of the phase
mismatch function with respect to $q$ and $\Omega$, equivalent to adopting the paraxial and quadratic dispersion approximations:
\begin{equation}
	\Delta(q,\Omega) =\Delta_0 l_c-\frac{q^2}{q_0^2}+\frac{\Omega^2}{\Omega_0^2} 
	\label{quadraticapprox}
\end{equation}
where $\Delta_0 = 2k_s - k_p$ is the collinear phase mismatch at degeneracy,  and 
\beqa
	q_0 &=& \sqrt{\frac{k_s}{l_c}} \, , \label{q0}\\
	\Omega_0&=&\sqrt{\frac{1}{k_s''l_c}}, \label{Omega0}
\eeqa
with $k_s = k_s(0)$, $k_s''=d^2 k_s/d\Omega^2|_0$. This expansion is strictly valid only for small $\Omega$ (close to degeneracy) and small ${q}$. For the remaining of this section we assume conditions of \emph{collinear} phase matching,  $\Delta_0 l_c$ $\approx$ 0, where the phase matching curve in the plane $ (q, \Omega)$ has the characteristic   hyperbolic geometry shown  in Fig.\ref{fig_spectrum}. The parameters
$q_0$, associated to spatial diffraction and $\Omega_0$, associated to group velocity dispersion (GVD),  define the typical variation scales of $|V(\vec{w})|^2$ along $q$ and $\Omega$,respectively. In the example of the 4 mm BBO crystal, their values are 
$q_0 \approx 5 \times 10^{-2} \mu {\mathrm m}^{- 1}$, $\Omega_0 \approx 0,76 \times 10^{14}$Hz. 
\\
ii) The second approximation consists in substituting the $\text{sinc}^2\left(\frac{\Delta(\w)}{2}\right)$ with a box function, with the same value of the indefinite integral
\begin{equation}
	\text{sinc}^2  ( \frac{\Delta(\w)}{2} )  \, \rightarrow \, \chi_{\alpha}( \frac{\Delta(\w)}{2} )  =\begin{cases}
	\frac{\pi}{\alpha} \hspace{0.4cm}  \frac{\Delta(\w)}{2}   \in(-\frac{\alpha}{2};\frac{\alpha}{2})\\
	0 \hspace{0.4cm}  \text{elsewhere}
	\end{cases}
	\label{sincapprox}
\end{equation}
which satisfies
$
\int{\text{sinc}^2(x)dx}= \int{\chi_{\alpha}(x)dx}=\pi
$. Here the parameter
$\alpha$ can be used in principle as a fitting parameter. Approximation (\ref{sincapprox}) seems very rough, but  it turned out surprisingly accurate: the rationale behind this result is that the sinc$^2$ has a sharp  peak where  $\Delta (\w) =0$, and in order to evaluate its integral in the 3D space it is more important to take into account the geometrical shape of the curve where its maximum lies rather than the detailed shape of the peak.  Figure \ref{figbox} compares the box function approximation to the true phase matching function in  the example of the BBO crystal. Here,  substantial deviations appear at large values of $\Omega$ and $q$ because of the failure of  the quadratic approximation for phase matching.  
\par 
Notice that if we also assume that  $|V(\vec{w})|^4$ can be approximated by the box function $\chi_{\alpha}^2(\vec{w})$, with the request that, this time, 
\begin{equation}
	\int{d\,x\, \chi^2_{\alpha}(x)} =\frac{\pi^2}{\alpha}= \int{d\,x\;\text{sinc}^4(x)}=\frac{2}{3}\pi  \, , 
\end{equation}
we  find the correct value of $\alpha =\frac{3}{2}  \pi.$
This also shows that, within these approximations,   the ratio
\begin{equation}
\frac{\int{d\vec{w}}|V(\vec{w})|^2}{\int{d\vec{w}}|V(\vec{w})|^4} 
\approx \frac{\int{d\vec{w}} \chi_{\alpha}(\vec{w})}{\int{d\vec{w}} \chi^2_{\alpha}(\vec{w})} =  \frac{\alpha}{\pi}= \frac{3}{2}  
	\label{ratio1}
\end{equation}
The box function approximation allow us to evaluate easily the integrals over the phase matching curves inside Eq. \eqref{k3PW}. 
In this evaluation, we assume that our model describes a measurement performed over a large but limited spectral bandwidth $\Omega\in(-\Omega_{max},\Omega_{max})$. For simplicity, here we do not pose limits to the spatial bandwidth (which will be instead done in the following section) . 
After some calculations we obtain the two different results depending on the detected bandwidth 
$\overline{\Omega}_{max}= \frac{\Omega_{max}} {\Omega_0}$:
\begin{itemize}
	\item \textbf{small bandwidth result} ($\overline{\Omega}_{max}<\sqrt{\alpha}$)
\begin{equation}
	\mathcal{K}= \frac{\alpha}{4} 
\sqrt { \frac{\alpha }    {\pi }  }   \;  
q_0^2\Omega_0 \sigma_p^2 \tau_p \left[   \frac{   \overline{\Omega}_{max} }  {\sqrt{\alpha}} 
+\frac{1} {3}     \left(\frac{   \overline{\Omega}_{max} }  {\sqrt{\alpha}}  \right)^3     \right],
	\label{k3d1}
\end{equation}
	\item \textbf{large bandwidth result} ($\overline{\Omega}_{max}>\sqrt{\alpha}$)
\begin{equation}
	\mathcal{K}=  \frac{\alpha}{2} 
\sqrt { \frac{\alpha }    {\pi }  }   \;  
q_0^2\Omega_0\sigma_p^2\tau_p 
 \left(\frac{\overline{\Omega}_{max}} {\sqrt{\alpha}}-\frac{1}{3} \right),
	\label{k3d2}
\end{equation}
\end{itemize}
First of all we observe that the condition on the bandwidth can be roughly reformulated as 
$\Omega_{max} $ being smaller or bigger than the characteristic GVD bandwidht $\Omega_0$ (since $\alpha$ is on the order of unity).
The small bandwidth case corresponds to the situation where the portion of PDC emission intercepted by the measurement lies within the central region of the phase matching curve (see  Fig.\ref{fig_spectrum}),  where the phase matching has no hyperbolic structure. 
Eqs. \eqref{k3d1}, \eqref{k3d2}  tell us that in both cases  the Schmidt number is  proportional to the number of modes contained 
in a unit volume of the phase matching region:   
\beq
\cappa \propto \frac {\pi q_0^2\Omega_0 \sigma_p^2 \tau_p }{8} = \frac {\pi q_0^2\Omega_0}{\delta \q_p^2 \, \delta \Omega_p} 
\eeq 
However, as the detected  bandwidth increases beyond the GVD bandwith $\Omega_0$,  the hyperbolic geometry of phase matching enters to play,  and the Schmidt number result of Eq.\eqref{k3d2}  shows  a linear increase with the bandwidth. 
\par 
We remind that the  analytical expressions \eqref{k3d1} and \eqref{k3d2} 
estimate the Schmidt number within the NPWPA and the quadratic approximation, expressed by the condition (\ref{NPWPA}) and \eqref{quadraticapprox}, respectively. 
In order to verify its validity, and at the same time  provide a numerical estimation of $\mathcal{K}$ in regions of the parameter space where the NPWPA does not hold, we performed a numerical evaluation of the general expression of $\mathcal{K}$  given by Eq.(\ref{kintegral}),(\ref{N}) and (\ref{k2}).  As it involves a 6-dimensional integral for the evaluation of $N$ and a 12-dimensional integral for the evaluation of $B$, the use of a Montecarlo integration is mandatory. We used the well known method of importance sampling \cite{Hammersley64}
with the aim of improving the efficiency of the Montecarlo algorithm by increasing the density of the sampled points
where the functions under the integrals  are larger. 
A natural choice has been to sample some of the  Fourier variables (namely the "pump" variables in Eq.\eqref{exactB}) according to Gaussian distributions coincident
with the Gaussian pump spectral amplitude \eqref{pumpspectrum}, which in the NPWPA represents the narrowest factor of the
 the biphoton amplitude in Eq.\eqref{psip}. 
The inplemented algorithm, very efficient for narrow specral pump profiles,  allows the evaluation of B and N 
even in the region where the the NPWP approximation  fails. No other substantial approximations are introduced, as the phase matching function is here evaluated  by means of the empirical Sellmeier formulas \cite{boeuf2000}. 
\begin{figure}[h]
\begin{center}
\includegraphics[width=8cm]{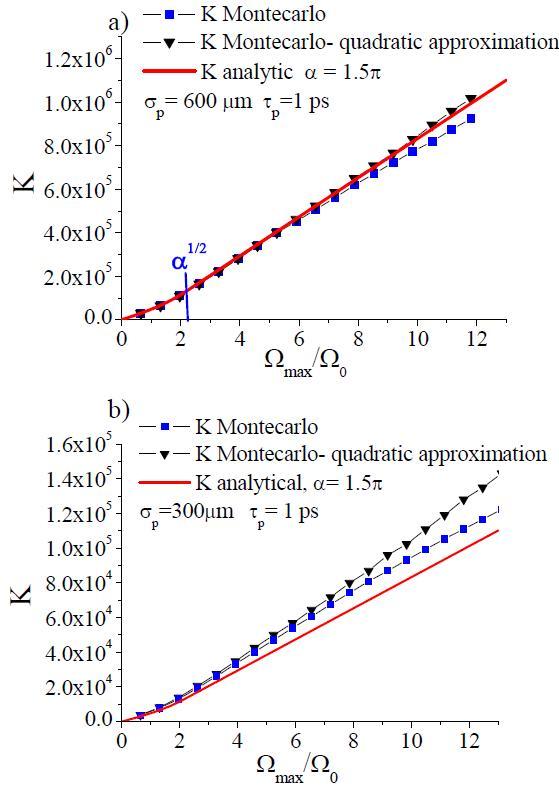}
\caption{Schmidt number results. Comparison between the analytic formula \eqref{k3d1}, \eqref{k3d2} (solid red line) and Montecarlo simulations, without any approximation (square) and with quadratic approximation (triangles). a) Pump parameters are within the NPWPA b) Focused pump, beyond the NPWPA. Collinear phase matching  ($\Delta_0 l_c =0$, $\theta_p = 22.934^o$),  $l_c=4$mm.  }
\label{result1}
\end{center}
\end{figure}
\par
Fig. \ref{result1} compares the analytic result \eqref{k3d1}, \eqref{k3d2}  with the Montecarlo numerics, performed both without approximation (squares) and with the quadratic approximation for phase matching (triangles). Part a) of the Figure is plotted for parameters of the pump within the NPWP approximation (althought very reasonable for an experimental realization) and shows an excellent agreement between the analytical curve  and the numerics, in the range of validity ogf the quadratic approximation. Indeed,  the analytic result follows very well a Montecarlo simulation performed with the quadratic approcximation, showing that the box function approximation  captures the basic geometrical properties of the phase matching function. In plot b) the pump beam is more focused and as expected the analytic result deviates from numerics because of the failure of the NPWP approximation. 
\begin{figure}[h]
\begin{center}
\includegraphics[width=8cm]{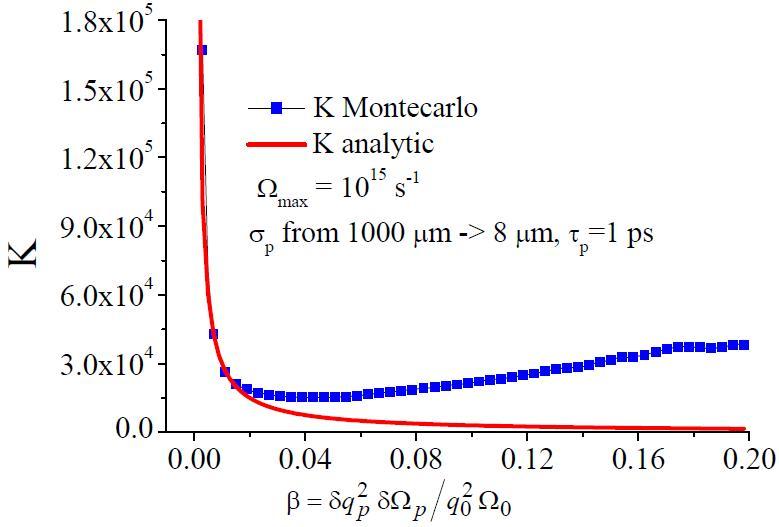}
\caption{Schmidt number as a function of $\beta= \delta q_p^2 \delta \Omega_p / q_0^2 \Omega_0$. The blue squares plot the result of a Montecarlo calculation,  without any approximation, and shows that $\cappa$ after reaching a minimum increases again almost linearly woth $\beta$.  The red solid line is the analytic result, decreasing as $1/\beta$ (see Eq.\eqref{k3d2}, valid only within the NPWPA (small $\beta$).  
Collinear phase matching  ($\Delta_0 l_c =0$, $\theta_p = 22.934^o$),  $l_c=4$mm.  }
\label{result2}
\end{center}
\end{figure}
\par 
Montecarlo calculations permit to obtain results also in the region of parameters beyond the NPWP approximation. Figure \ref{result2} plots the Schmidt number as a function of the parameter $\beta= \delta q_p^2 \delta \Omega_p / q_0^2 \Omega_0$. The NPWPA approximation is valid only for $\beta \ll 1$, i.e.   when the widths $\delta \Omega_p$, $\delta q_p$ 
of the pump Fourier profile are much smaller than the characteristic scales of variation of phase matching $\Omega_0$, $q_0$, respectively \footnote{Actually the limits of validity of the NPWPA expressed by \eqref{taup} and \eqref{sigmap} are typically much more restrictive than $\beta \ll 1$. For example for the BBO crystal here considered  $ \delta q_p < q_0$ implies roughly 
$\sigma_p > 2  l_{\mathrm diff} = 2/q_0 \approx 40\mu$m, while $\delta \Omega_p < \Omega_0$ implies 
$ \tau_p > 2 \tau_{\mathrm GVD} = 2/ \Omega_0 \approx 26$fs.}. 
The Montecarlo result shows a decrease of the Schmidt number as $1/\beta$ for $\beta \ll 1$, as predicted by the analytic result \eqref{k3d2} in the NPWPA (red solid line in the Figure). However, after reaching a minimum the Schmidt number increases again almost linearly with $\beta$. This behaviour is very similar to the one predicted in a purely spatial model of PDC in Ref.\cite{law2004} and can be understood as follows: for a broad pump, when the NPWPA is valid, the width of the correlation is determined by the pump Fourier profile, and the number of spatio temporal modes can be estimated as in formula \eqref{intgeom} as being proportional to the volume of phase matching divided  by the pump spectral volume 
$\cappa \propto q_0^2 \Omega_0/ \delta q_p^2 \delta \Omega_p = \frac{1}{\beta} $.  For a very focused pump,instead, the phase matching function in the $\q$ direction has a smaller scale of variation than the pump Fourier spatial profile, so that the width of correlation is rather determined by the characteristic width $\q_0$ of phase matching, and we can suppose that the number of modes is now  
$\cappa \propto   \delta q_p^2  / q_0^2 \propto \beta $. 
\section{Factorability of the Schmidt number in its temporal and spatial components}
In the literature the Schmidt dimensionality of twin photons  is often  calculated within models of PDC  restricted to the spatial or the temporal domain (see \cite{law2000,mikhailova2008,law2004,exter2006}. 
The non-factorable character of the spatio-temporal correlation demonstrated in \cite{gatti2009,caspani2010}  suggests us that the full 3D spatio-temporal Schmidt number is not trivially given by the product of the spatial 2D and the temporal 1D Scmhidt numbers. In this section  we would like to understand  this point.
\par 
To this end we consider models for PDC in lower dimensionalities, and follow the same procedure outlined in the previous sections to calculate the Schmidt number. The purely spatial $2D$ model is obtained by neglecting the temporal coordinate and setting $\Omega=0$. Similarly the purely temporal $1D$ model neglects the spatial coordinates and sets $\q=0$. 
The starting point of the analysis is in both cases the general integral formula for the Schmidt number \eqref{kintegral} where 
we have now to interpret the Fourier coordinates as: 
\begin{equation}
	\vec{w}\;=\;\left\{ \begin{array}{rcc}
	\Omega &\in \mathcal{R} &in \;\; 1D \\
	\vec{q} &\in \mathcal{R}^2 &in \;\; 2D \\
	\vec{q},\Omega &\in \mathcal{R}^3 &in \;\; 3D 
	\end{array} \right.
\end{equation}
Simlarly, in the expression involving the coordinates in the direct space:   
\begin{equation}
	\vec{\xi}\;=\;\left\{ \begin{array}{rcc}
	t &\in \mathcal{R} &in \;\; 1D \\
	\vec{x} &\in \mathcal{R}^2 &in \;\; 2D \\
	\vec{x},t &\in \mathcal{R}^3 &in \;\; 3D 
	\end{array} \right.
\end{equation}
For example, by introducing the NPWP approximation in the various models ( clearly NWPA  in the spatial model means that the pump has a broad waist, while the temporal model assumes a long enough pulse duration), we obtain the NPWPA expression for the Schmidt number in an arbitrary D-dimensional model:
\begin{equation}
\mathcal{K}=
\frac{  \left[ \int{d\vec{\xi}_p}|\mathcal{A}_p(\vec{\xi}_p)|^2 \right]^2}
{\int{d\vec{\xi}_p}|\mathcal{A}_p(\vec{\xi}_p)|^4} 
\; 
\frac{ \left[ \int{d\vec{w}}|V(\vec{w})|^2 \right]^2}
{(2\pi)^D   \, \int{d\vec{w}}|V(\vec{w})|^4} \, . 
\label{kgenPW}
\end{equation}
\subsection {Spatial Schmidt number $\cappa_{2D} $} 
By performing calculations similar to those reported for the 3D model, we derive an expression for the Schmidt number in the purely spatial case, valid within  the NPWPA and the quadratic approximation for phase matching. The latter one corresponds to approximating the phase matching function as 
\beq
\Delta_{2D} (\q) =k_s (\q) +k_s(-\q) -k_p \approx \frac {q^2}   {q_0^2} 
\label{Delta2d} 
\eeq 
In the $2D$  case the result depends on  the spatial bandwidht $\overline{q}_{max} = q_{max}/q_0$ intercepted by the measurement: 
\begin{mathletters}
\beqa 
	\cappa_{2D} =&  \frac{3}{8} \pi \sigma_p^2 q_0^2 \frac{ \overline{q}_{max}^2 }{\alpha} \quad  & \overline{q}_{max} < \sqrt{\alpha} 
	\label{k2d1}   \\
	\mathcal{K}_{2D}=& \frac{3}{8} \pi \sigma_p^2 q_0^2  	\quad  & \overline{q}_{max} > \sqrt{\alpha} 
	\label{k2d2}
\eeqa
\end{mathletters}
where we remind that  $\alpha \approx 1. 5 \pi$. 
\begin{figure}[h]
\begin{center}
\includegraphics[width=8cm, keepaspectratio]{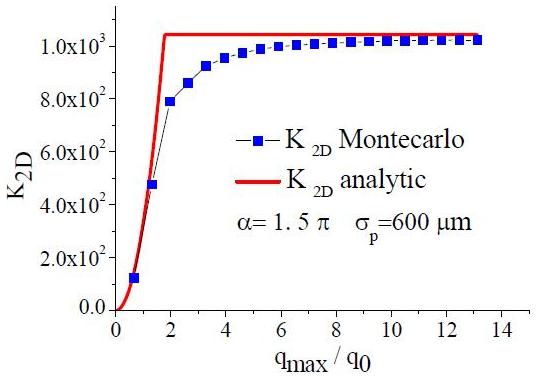}
\caption{2D spatial Schmidt number $\mathcal{K}_{2D}$ as a function of the collected spatial bandwidth  ${q}_{max}$. The solid red line is the analytic result \eqref{k2d1}, \eqref{k2d2}, the squares plot the  Monte Carlo numeric result. The waist of the pump beam is $\sigma_p$ = $600 \mu$m.}
\label{result2d}
\end{center}
\end{figure}
In Fig.\ref{result2d} this curve  is compared with an exact Montecarlo calculation performed in the $2D$ model. Beyond noticing that the two results agree qualitatively, we remark that, differently from the 3D case, the 2D Schmidt number saturates to the maximum value $\mathcal{K}_{2D max}= \frac{3}{8}\pi\sigma_p^2 q_0^2 = \frac{3}{2}\pi q_0^2/ \delta q_p^2 $. This behaviour can be explained with the help of the geometrical interpretation (\ref{intgeom}), valid in the NPWPA, which evaluates 
 the Schmidt number as the ratio between the volume of the phase matching region and the correlation volume.  In the 2D case  phase matching is described by Eq. \eqref{Delta2d},  so that in the $(q_x, q_y)$ plane phase matching occurs within a circle  of area  $\approx \pi \q_0^2$.  For increasing $q_{max}$, the  PDC photons are detected in increasing large circular regions, so that  the Schmidt number increases quadratically with $q_{max}$ until the border of the phase matching region $q_{max}= q_0 \sqrt{\alpha}$ is reached. 
\subsection {Temporal  Schmidt number $\cappa_{1D} $} 
We now consider the purely temporal model of PDC, by setting $\q=0$. We  perform again analytic calculation in the NPWPA and the quadratic approximation for phase matching, that in the 1D case reads
\beq
\Delta_{1D} (\Omega) =k_s (\Omega) +k_s(-\Omega) -k_p \approx \frac {\Omega^2}   {\Omega_0^2} 
\label{Delta1d} 
\eeq 
The analytic expression for $\cappa_{1D}$ in these limits, obtained by using the box function approximation,  depends on the  collected temporal bandwidth $\overline{\Omega}_{max} = \Omega_{max}/\Omega_0$: 
\begin{mathletters}
\beqa 
	\cappa_{1D} =&  \sqrt{ \frac{\alpha}{\pi}} 
          \tau_p \Omega_0 \frac{ \overline{\Omega}_{max} } {\sqrt{\alpha}} \quad  & \overline{\Omega}_{max} < \sqrt{\alpha} 
	\label{k1d1}   \\
	\mathcal{K}_{1 D}=& \sqrt{ \frac{\alpha}{\pi}} 
			         \tau_p \Omega_0 
 					\quad  & \overline{\Omega}_{max} > \sqrt{\alpha} 
	\label{k1d2}
\eeqa
\end{mathletters}
\begin{figure}[htbp]
\begin{center}
\includegraphics[width=8cm, keepaspectratio]{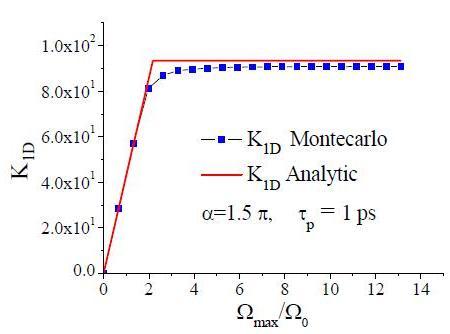}
\caption{1D temporal Schmidt number $\cappa_{1D}$ as a function of the maximum temporal frequency $\Omega_{max}$. The solid line shows the analytic result (within the NPWPA and quadratic approximation), the squares plot the numeric exact result. Pump time duration $\tau_p = 1 $ps.}
\label{result1d}
\end{center}
\end{figure}
Figure \ref{result1d} plots this analytical result toghether  with the exact Montecarlo 1D calculation. Also in this case, similarly to the 2D case, the Schmidt number saturates to the maximum value 
$\mathcal{K}_{1D max} \approx \tau_p \Omega_0$, because phase matching occurs only inside an interval of size $\approx \Omega_0$, so that 
 by increasing  $\Omega_{max}$ beyond the critical value $\sqrt{\alpha} \Omega_0$ , we begin to consider regions where there is no phase matching, which do not contribute to the integral.
\subsection{Comparison} 
\begin{figure}
\begin{center}
\includegraphics[width=8cm, keepaspectratio]{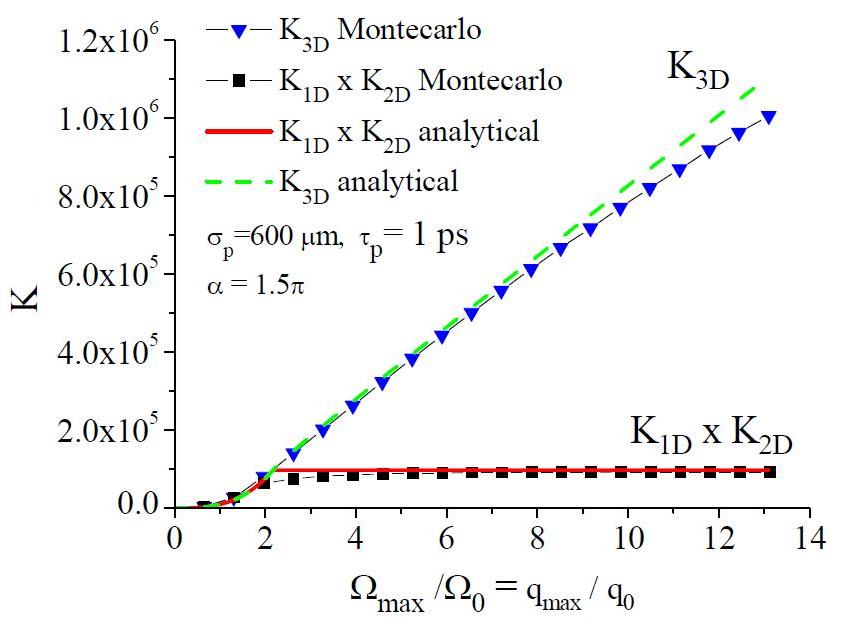}
\caption{Comparison between the Schmidt number $\mathcal{K}$ calculated in the full spatio-temporal model and the product of the  Schmidt numbers $\mathcal{K}_{1D} \times \mathcal{K}_{2D}$,  calculated in the purely 1D temporal and 2D spatial models. The abscissa is the collected temporal   bandwidth, set equal to the spatial bandwidth in  normalized units. Lines plot analytic results, symbols provide the Montecarlo results.} 
\label{result3}
\end{center}
\end{figure}

We are now in conditions of comparing  the results obtained in the models of various dimensionalities. 
To this end, we have to reformulate slightly the 3D problem. In Section\ref{Sec:3D} we calculated $\cappa$ as a function of the collected temporal bandwidth by assuming that no selection on the spatial bandwith was performed, i.e.  $\overline{q}_{max}=\infty$. This is a possible correct choice to present results, but in order  to have a clean comparison with the 2D and 1D models, we need also to limit the detected spatial bandwidth. The simplest possibility is to set  $\overline{q}_{max}= \overline{\Omega}_{max}$. This choice is justified by the fact that in the quadratic approximation, phase matching is realized along the lines $
	\frac{q}{q_0} = \pm \frac{\Omega}{\Omega_0}$ 
so that by increasing simultaneously the spatial and temporal bandwidth $\frac{q_{max}}{q_0}=\frac{\Omega_{max}}{\Omega_0}$ one  follows the phase matching curve. \\
With this in mind, analytical calculations in the PWPA, quadratic approximation for phase matching and box function approximation can be performed. The result for the $3D$ Schmidt number is plotted in Fig.\ref{result3} (dashed line) toghether with the Montecarlo exact result (triangles). In the same figure we compare these 3D results with the product of the Schmidt numbers obtained in the models with lower dimensionalities i.e. $\cappa_{2d}\,  \times \, \cappa_{1D}$.   
From these plots it clearly emerges that  the factorizability holds only when the detected bandwidth is  small, i.e. when both $q_{max}$ and $\Omega_{max}$ lie within the phase matching bandwiths $q_0$, $\Omega_0$, respectively. However, as the detected bandwidth gets larger, the result in the fully $3D$ spatiotemporal model grows linearly with the detected  bandwith, and  diverges clearly from the product  $\cappa_{2d}\,  \times \, \cappa_{1D}$ which saturats to a fixed value $\propto q_0^2  \Omega_0$.
This result can be easily understood with the help of the geometrical interpretation of the Schmidt number  as the number of entangled modes contained in the phase matching region:  close to the degeneracy and to the collinear emission, the phase matching region can be seen as a spherical region, which obviously factorizes in the spatial and temporal components, so that the number of spatio-temporal modes is trivially the product of the numbers of spatial and temporal modes times. 
 Conversely, if the collected bandwith is large enough, than  the biconical,  non-factorizable geometry of phase matching comes into play,  so that a full 3D model has to be used to correctly compute the number of spatio-temporal modes. 
\section{Entanglement without phase matching}
\label{sec_noncoll}
\begin{figure}
\begin{center}
\includegraphics[width=7cm, keepaspectratio]{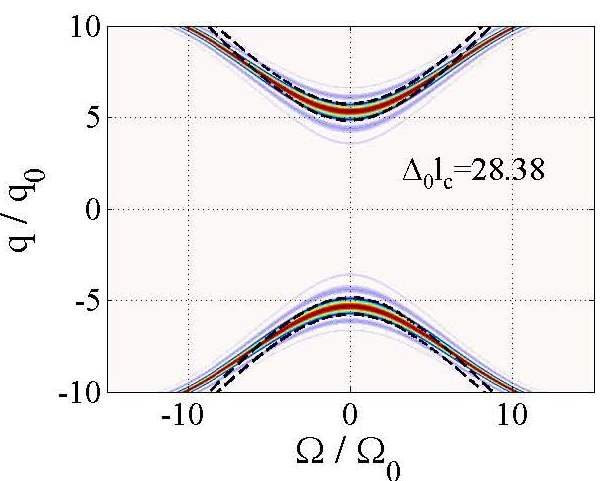}
\caption{Colormap plot of $|V(\vec{q},\Omega)|^2 $ in the non-collinear phase matching case, $\Delta_0 l_c > 0$,($\Delta_0 l_c = 23.38$). The dashed lines are the boundaries of the box function, where $\Delta(\vec{q},\Omega)= \pm\alpha$ } 
\label{PMnoncollinear}
\end{center}
\end{figure}

Up to now we considered the case of collinear phase matching ($\Delta_0 l_c= 0$), characterized by the fact that exact phase matching $\Delta (\w) =0 $ can be realized for any  value of the transverse wave-vector  of the  photon pair, such that 
 $q/q_0= \pm \Omega/\Omega_0 $. 
However, if the crystal is tuned away from the collinear conditions  ($\Delta_0 l_c \ne  0$), there exist regions of the $(\q, \Omega)$ space where phase matching does not occur at all. In these regions,  the probability of emission of photon pairs is low, although not zero. \\
In particular,  we focus on the case of non-collinear phase matching ($\Delta_0 l_c>0$). Fig.\ref{PMnoncollinear} shows the typical  behaviour of the phase matching function ${\rm sinc}^2 (\Delta (\w) /2)$ . From the quadratic expansion of the phase mismatch 
\begin{equation}
	\Delta(q,\Omega)\approx \Delta_0 l_c -\frac{q^2}{q_0^2}+\frac{\Omega^2}{\Omega_0^2},
\end{equation}   
we immediately realize that phase matching does not occur for modes close to collinear emission, namely having $|\q| < \sqrt{  \Delta_0 l_c } $ 
\par
\begin{figure}[h]
\begin{center}
\includegraphics[width=8cm, keepaspectratio]{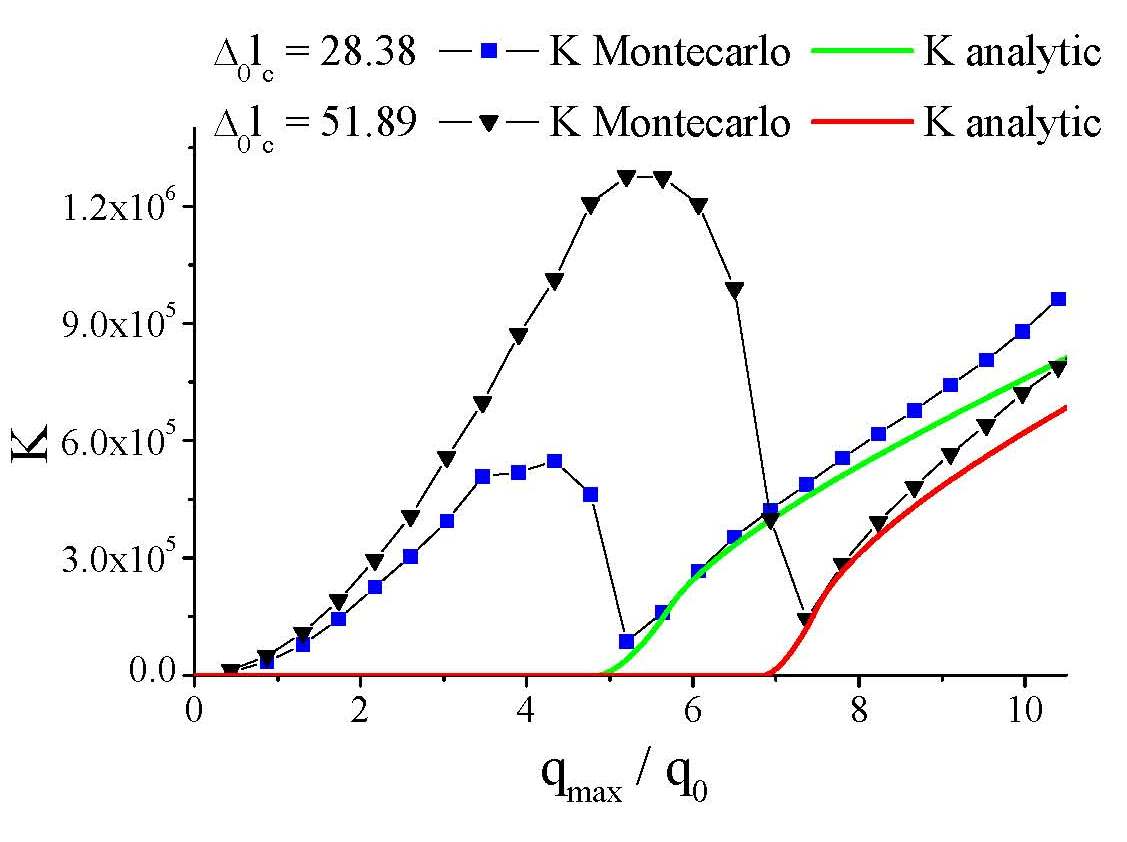}
\caption{Schmidt number $\mathcal{K}$ in the non-collinear phase matching case ,  as a function of the collected spatial bandwith $q_{max}$,  normalized to the diffraction bandwith $q_0$. Parameters are $\sigma_p = 600 \mu$m, $\tau_p =1$ps}
\label{result5}
\end{center}
\end{figure}

\begin{figure}[h]
\begin{center}
\includegraphics[width=7.5cm, keepaspectratio]{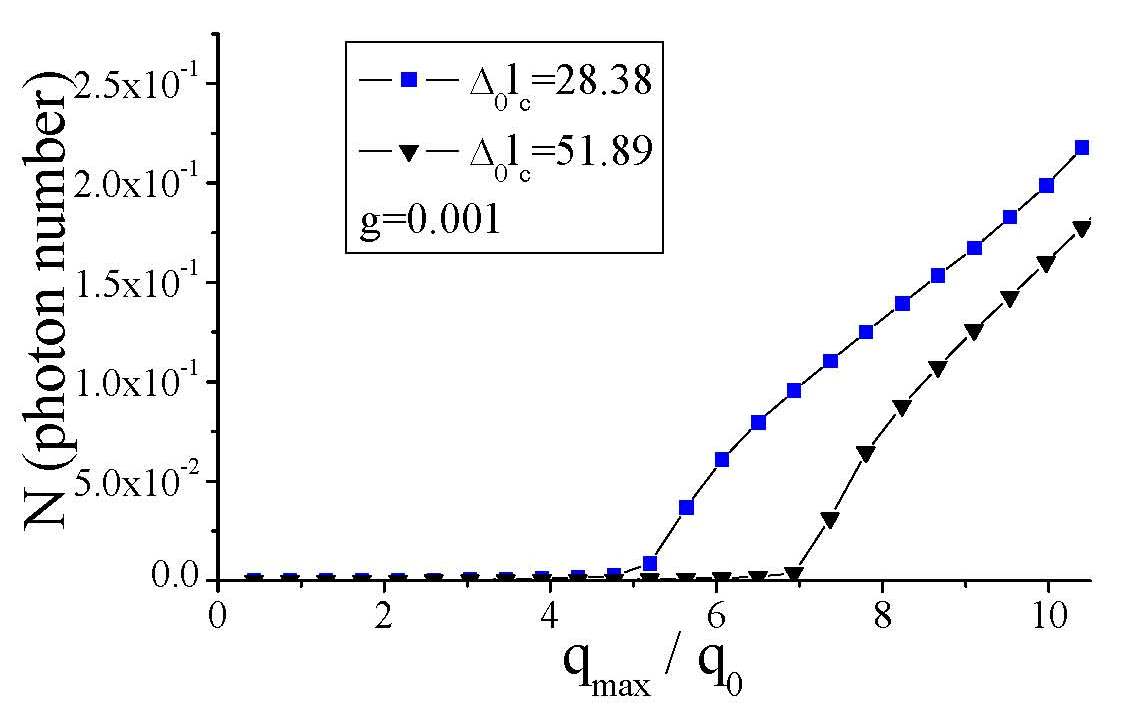}
\caption{Number of photons $N$ in the non-collinear case, as a function of $\overline{q}_{max}$. Parameters are $\sigma_= 600\mu$m, $\tau_p$ $=$ $1$ ps, $g=0.001$.}
\label{N2}
\end{center}
\end{figure}

 
We have calculated  the 3D  spatio-temporal Schmidt number in these conditions:  Fig. \ref{result5} plots the result as a function of the collected spatial bandwidth $q_{max}$. We find that for small bandwidths,  such that the collected photons are not phase matched, the degree of entanglement is very high , actually larger than in the region where phase matching is realized. This result is  apparently paradoxical, because in the absence of phase matching the probability of emission of a photon pair is very low and one could infer that the state should be very close  to the separable vacuum state.
For comparison Fig.\ref{N2} displays the corresponding mean number of photons, which as expected, is indeed very low where there is no phase matching.  
\par
However, in order to understand the  results of Fig. \ref{result5} we have to remind that we are studying the degree of entanglement of the state \eqref{state2}, conditioned to the detection of a photon pair.  This means that our  calculation of the Schmidt number  has lost track of the presence of a large  vacuum contribution to the original PDC state, and the result has to be interpreted as { \em photon pairs are emitted with very low probabilty, however,  when a pair is detected it is highly entangled}.\\ 
The point to understand is therefore why the non-phase matched photon pairs are more entangled than the phase matched ones. 
The Schmidt number $\mathcal{K}$ = $1/\sum_j{\lambda_j}^2$ provides an estimate of the number of significant eigenvalues of the Schmidt decomposition, i.e. the number of entangled eigenmodes that participate to the modal decomposition. In the region where no phase-matching occurs (for $\overline{q}_{max}< \sqrt{\Delta_0 l_c}$), $\mathcal{K}$ can be in practice very large, since there is no mechanism for modal selection and  all the modes in the collected bandwidth participate equally to the PDC process even though with a very low occupation probability. By contrast, when increasing $\overline{q}_{max}$ towards
$\sqrt{\Delta_0 l_c}$ and entering the phase-matching region, a strongly reduced number of phase-matched spatio-temporal modes  contribute to the Schmidt decomposition, the few which are close to satisfy the phase-matching  condition $\overline{q}$ = $\sqrt{\Delta_0 l_c}$, $\Omega$ = 0. The number of significant eigenmodes is therefore reduced  because phase matching operates a selection of the spatio-temporal modes that efficiently participate to the entanglement of the state. 
In other terms, the phase-matching mechanism favors only a small number of  modes, reducing thereby drastically the dimensionality of the entangled state. By increasing $\overline{q}_{max}$ above $\sqrt{\Delta_0 l_c}$ the Schmidt number $\mathcal{K}$ starts again to increase, according to the geometrical interpretation of the Schmidt number as being proportional again to the phase matching volume
\section{Conclusions}
In this work we have calculated the Schmidt dimensionality of the two-photon state generated by PDC in the ultra-low gain regime. Results have been produced with different degrees of approximation: Montecarlo calculations  allow us to obtain the Schmidt number without relevant approximations, while when the pump beam is broad enough  (NPWPA) we have demonstrated a trasparent geometrical interpretation of the Schmidt number, which can be seen  as the number of entangled modes contained in the region where phase matching occurs.
\par 
The same calculations have been performed in models restricted to the purely spatial or temporal degrees of freedom of biphotons.  A remarkable result  demonstrated is that,  when the collected spatio-temporal bandwidth is large enough, the  Schmidt dimensionality of the full spatio-temporal state cannot be trivially reduced to the product of the Schmidt numbers characterizing the entanglement in lower dimensions. Therefore, obtaining the Schmidt number in the full 3D model is not a mere exercise of calculus:  in order to correctly characterize the  entanglement of twin photons it is necessary to consider simoultaneously their spatial and temporal degrees of freedom. 
This result is a clear consequence of the non-factorability of the state in apace and time, and mirrors the findings described in Refs. \cite{gatti2009}, \cite{caspani2010} where the spatio-temporal correlation of the biphoton state was shown to have a non factorable  X-shaped geometry. 
\par
The non-factorability has been demonstrated in this work only in the NPWPA. Question is still open whether in the opposite limit, that is, for a very focused and short pump pulse,  the Schmidt number keeps the non-factorable character, and it is obviously linked to the more general question whether the state appears to be factorable in space and time in these conditions. 
 We however remark that 
reaching this limit  is in practice very demanding, because it requires that the pump Fourier profile 
{\bf at the exit face of the crystal} is much broader than the width of the phase matching function. For a few millimeter crystal,  this implies a pump waist on the order of tens of microns, and a pulse duration as short as few tens of femtosecond.  While the first condition could be in principle achieved by strongly focusing at the end face of the crystal, the second is  more demanding because of dispersion occuring inside the nonlinear material. 
\par
An intriguing finding is that in the absence of  phase matching, where the probability of emission of photon pair is very low, the Schmidt dimensionality of the state is huge, actually larger than in the regions  where phase matching occurs. This counter-intuitive finding has been explained throught the mode selection mechanism performed by phase matching, which reduces the available number of spatio temporal modes. However, in evaluating this result and its usefulness for applications, one has to remember that the Schmidt number here analysed does not refer to the  full PDC state, but to the state conditioned to the detection of a photon pair. 
\acknowledgments{We acknowledge support from the Fet Open programme of the EC within the grant 221906 HIDEAS {\em High Dimensional Entangled Systems} }
\appendix
\section{Partial density matrix}
\label{A2}
In order to calculate the partial trace of the density matrix \eqref{rho} it is enough to consider the vacuum state $\ket{0}_2$ of the "idler"  subsystem 2, plus the continuous set of  1 photon states 
\beq
\left\{ A^{\dagger}_2 (\w_2) \ket{0}_2 \right\}
\label{set} 
\eeq
It is convenient to write the density matrix \eqref{rho} as 
\beq
\rho= \frac{1}{N} M^{\dagger} \ket{0} \bra{0} M
\eeq
where $M$ is the operator that annihilates a photon pair in any spatiotemporal mode (weighted by $\psi$)  
\beq
M= \int d\w_1 \int d\w_2 \psi^* (\w_1,\w_2) A_1(\w_1) A_2(\w_2) 
\eeq
We then have
\begin{widetext} 
\beqa
\rho_1 &=& {\mathrm Tr}_2 \{ {\rho} \}  \nn \\
&=& \frac{1}{N} \left \{    \zerodue M^\dagger \ket{0}_2  \ket{0}_1 \zerouno  \zerodue M  \ket{0}_2 
	+ \int d\w_3 \zerodue A_2(\w_3) M^{\dagger} \ket{0}_2  \ket{0}_1   \zerouno   \zerodue M A_2^{\dagger} (\w_3) \ket{0}_2  
\right\}  \label{l1} \\
&=&  \frac{1}{N} \left \{ \int d\w_3 \int d\w_1 \int d\w_2 \int d\w_1' \int d\w_2' \psi(\w_1, \w_2) \psi^*(\w_1', \w_2')
	\right.  \nn \\  
	&\times& \left. \phantom{\int} \! \! \! \zerodue A_2(\w_3) A_2^{\dagger} (\w_2) \ket{0}_2 A_1^{\dagger} (\w_1) \ket{0}_1 
	\zerouno A_1(\w_1') \zerodue A_2(\w_2) A_2^{\dagger} (\w_3) \ket{0}_2 \right\}  \label{l2} \\
&=& \frac{1}{N} \left \{ \int d\w_1 \int d\w_2 \int d\w_1'  \psi(\w_1, \w_2) \psi^*(\w_1', \w_2) 
	A_1^{\dagger} (\w_1) \ket{0}_1 \zerouno A_1(\w_1') \right\} \label{l3} \\
&=& \frac{1}{N} \left \{ \int d\w_1  \int d\w_1'  G(\w_1', \w_1) A_1^{\dagger} (\w_1) \ket{0}_1 \zerouno A_1 (\w_1') \right\} 
\label{l4}
\eeqa
\end{widetext}
where in passing from \eqref{l2} to \eqref{l3} we used the relation
\beq
\zerodue A_2(\w_2) A_2^{\dagger} (\w_2') \ket{0}_2 = \delta (\w_2-\w_2')
\eeq
which comes directly from the commutation rules of bosonic operators. 
\section{Derivation of formula (\ref{k2PW})}
\label{AppendixB}
We wish here to simplify the general formula \eqref{kintegral} by exploiting the NPWPA introduced in Section \ref{sec:NPWPA} 
(see Eq.\eqref{NPWPA}).
We rewrite formula \eqref{kintegral} as 
\beq
\cappa = \frac{N'^2}{B'}
\eeq
\begin{widetext}
with 
\beqa
B'  &=& \int{d\vec{w}_1}\int{d\vec{w}_2}\int{d\vec{w}'_1}\int{d\vec{w}'_2} 
\tilde{\mathcal{A}_p}(\vec{w}_1+\vec{w}_2)\tilde{\mathcal{A}_p} (\vec{w}'_1+\vec{w}'_2) 
\tilde{\mathcal{A}_p}^*(\vec{w}'_1+\vec{w}_2)      \nn \\
 & &
\tilde{\mathcal{A}_p}^*(\vec{w}'_1+\vec{w}_2)      \tilde{\mathcal{A}_p}^*(\vec{w}_1+\vec{w}'_2)\cdot  
V(\vec{w}_1,\vec{w}_2)V(\vec{w}'_1,\vec{w}'_2)V^*(\vec{w}'_1,\vec{w}_2)V^*(\vec{w}_1,\vec{w}'_2)\, ;
\label{Bp}   \\
	N' &=& \int{d\vec{w}_1 \, d\vec{w}_2  
|\tilde{\mathcal{A}}_p(\vec{w}_1+\vec{w}_2)|^2|V(\vec{w}_1,\vec{w}_2)|^2}.
\label{Np}
\eeqa
\end{widetext}
where we inserted the explicit expression \eqref{psip} for the biphoton amplitude (without a costant coefficient). Here the function $V$ depends on phase matching and is given by Eq.\eqref{V}, while $\Aptilde$ is the Fourier profile of the pump. 
We start by simplifying the integral $B'$ in Eq.\eqref{Bp}, by introducing the change of variables
\begin{widetext}
\begin{equation}
	\left( \vec{w}_1, 
	\vec{w}'_1
	\vec{w}_2, 
	\vec{w}'_2 \right) \rightarrow 
	\left( \vec{w}_1, 
	\vec{\delta} = \vec{w}_1 -\vec{w}'_1, 
	\vec{w}_p=\vec{w}_1+\vec{w}_2, 
	\vec{w}'_p=\vec{w}'_1+\vec{w}'_2
	\right)
\end{equation}
%
With this  change, \emph{B'} becomes
\begin{align}
	B' =\int d\vec{w}_p \, &d\vec{w}'_p \, d\vec{\delta} \, d\vec{w}_1 \, \tilde{\mathcal{A}_p}(\vec{w}_p) \tilde{\mathcal{A}_p}(\vec{w}'_p) \tilde{\mathcal{A}_p}^*(\vec{w}_p-\vec{\delta})\tilde{\mathcal{A}_p}^*(\vec{w}'_p+\vec{\delta})
V(\vec{w}_1,-\vec{w}_1+\vec{w}_p)  \cr
& V(\vec{w}_1-\vec{\delta},-\vec{w}_1+\vec{\delta}+\vec{w}'_p)
V^*(\vec{w}_1-\vec{\delta},-\vec{w}_1+\vec{w}_p)V^*(\vec{w}_1,-\vec{w}_1+\vec{\delta}+\vec{w}'_p).
\label{exactB}
\end{align}
\end{widetext}
In this expression the variables $\vec{w}_p$ and $\vec{w}'_p$ die on the scale of the inverse of the  pump waist/duration, i.e. $(2/\sigma_p,2/\tau_p)$. Because of the presence of the terms $\mathcal{A}^*_p(\vec{w}'_1-\vec{w}_1+\vec{w}_p)$ and $\mathcal{A}_p^*(\vec{w}_1-\vec{w}'_1+\vec{w}'_p)$, also the variable $\delta$ $=$ $\vec{w}_1-\vec{w}'_1$ dies out on the same scale. We can then make use of the NPWP approximation, which amounts to substituting 
\beqa
V(\vec{w}_1,-\vec{w}_1+\vec{w}_p)  &\approx& V(\vec{w}_1-\vec{\delta},-\vec{w}_1+  \vec{\delta}+\vec{w}'_p) \nn\\
&\approx &
V(\vec{w}_1,-\vec{w}_1)  : = V(\w_1) 
\eeqa
where we took into account that all the pump variables $\w_p$, $\vec{\delta}$, $\vec{\delta}+\vec{w}'_p$ dies out on the fast scale 
of the inverse of the pump waist/duration, over which the function $V$ remains constant. Similarly 
\beqa
V^*(\vec{w}_1-\vec{\delta} ,-\vec{w}_1+\vec{w}_p)  &\approx& V^*(\vec{w}_1, -\vec{w}_1 +\vec{\delta} +\vec{w}'_p) \nn\\
&\approx &
V^*(\vec{w}_1,-\vec{w}_1)  : = V^*(\w_1) 
\eeqa
Within the NPWP approximation we hence obtain a new expression for $B'$, which reads:
\begin{widetext}
\beqa
B' &=&\int{d\vec{w}_p \, d\vec{w}'_p \, d\vec{\delta}}   \tilde{\mathcal{A}_p}(\vec{w}_p)\tilde{\mathcal{A}_p}(\vec{w}'_p)\tilde{\mathcal{A}_p}^*(\vec{w}_p-\vec{\delta})\tilde{\mathcal{A}_p}^*(\vec{w}'_p+\vec{\delta}) 
\times \int{d\vec{w}_1}|V(\vec{w}_1)|^4.
\label{approxB}
\eeqa
The integral over the pump variables can be further simplified by noting that it involves two convolutions:
\begin{equation}
	\int{d\vec{w}_p  \tilde{\mathcal{A}_p}(\vec{w}_p) \, \tilde{\mathcal{A}}^*_p(\vec{w}_p \pm \vec{ \delta})}=
\int{d\vec{\xi}_p} |\mathcal{A}_p(\vec{\xi}_p)|^2 {\rm e}^{-\pm i\vec{\xi}_p \cdot \vec{\delta}}  =  (2\pi)^{\frac{3}{2}} \left[\mathcal{F}(|\mathcal{A}_p|^2)\right](\pm \vec{\delta}),
\end{equation}
where the symbol $\mathcal{F} (f) $ denotes the Fourier transform of the function f . 
By performing also the integration over $\vec{\delta}$ we obtain
\beqa
\int{d\vec{\delta}}\left[\mathcal{F}(|\mathcal{A}_p|^2)\right](\vec{\delta})\left[\mathcal{F}(|\mathcal{A}_p|^2)\right](-\vec{\delta}) =
\int{d\vec{\xi}_p}|\mathcal{A}_p(\vec{\xi}_p)|^4,
\eeqa
\end{widetext}
where we used the Plancherel theorem $\int{d^3\vec{w}|\tilde{f}(\vec{w})|}^2=\int{d^3\xi|f(\xi)|}^2$. This leads to:
\begin{equation}
	B'=(2\pi)^3\left[\int{d\vec{\xi}_p}|\mathcal{A}_p(\xi_p)|^4\right]\left[\int{d\vec{w}}|V_{pw}(\vec{w})|^4\right].
\end{equation}
In order to complete the Schmidt number calculation we also have to evaluate $N'$, proportional to the mean photon number.  
With the usual change of variables $(\vec{w}_1,\vec{w}_2)\rightarrow(\vec{w}_p=\vec{w}_1+\vec{w}_2,\vec{w}_1)$, equation \eqref{Np}  becomes: 
\begin{equation}
N' =\int{d\vec{w}_p\;|\tilde{\mathcal{A}}_p(\vec{w}_p)|^2}\int{d\vec{w}_1\;|V(\vec{w}_1,-\vec{w}_1+\vec{w}_p)|^2}.
\end{equation}
In the NPWP limit (see Eq. \ref{NPWPA}) we get
\begin{equation}
N'=\int{d\vec{w}_p|\tilde{\mathcal{A}}_p(\vec{w}_p)|^2\int{d\vec{w}}|V(\vec{w})|^2}\, ,
\end{equation}
which, using the Plancherel theorem in the first integral,  can be also written as
\begin{equation}
N'=\int{d\vec{\xi}_p|\tilde{\mathcal{A}}_p(\vec{\xi}_p)|^2\int{d\vec{w}}|V(\vec{w})|^2}.
\end{equation}
In NPWP limit the Schmidt number takes therefore the symplified form
\begin{equation}
\mathcal{K}=\frac{N^2}{B}=
\frac{  \left[ \int{d\vec{\xi}_p}|\mathcal{A}_p(\vec{\xi}_p)|^2 \right]^2}
{\int{d\vec{\xi}_p}|\mathcal{A}_p(\vec{\xi}_p)|^4} 
\; 
\frac{ \left[ \int{d\vec{w}}|V(\vec{w})|^2 \right]^2}
{(2\pi)^3   \, \int{d\vec{w}}|V(\vec{w})|^4}
\label{k1PW}
\end{equation}
where the integrals now  factorize into the pump and signal degrees of freedom. 

\end{document}